\documentclass[letter,twocolumn]{jpsj3}

\allowdisplaybreaks[1]

\usepackage{txfonts}
\usepackage{cite}
\usepackage{graphicx}
\usepackage{bm}
\usepackage{braket}
\usepackage{cases}
\usepackage{mathrsfs}
\usepackage{amsmath}

\usepackage{color}
\usepackage{soul}
\usepackage{comment}

\usepackage{ulem}

\usepackage{float}

\title{Current-Induced Spin-Wave Doppler Shift in Antiferromagnets}
\date{\today}
\author{Jotaro J.~Nakane and Hiroshi Kohno}
\inst{Department of Physics, Nagoya University, Nagoya 464-8602, Japan} 

\abst{

We theoretically study the spin dynamics in antiferromagnets (AFs) 
under the influence of an electric current. 
We identify two different sources of spin-transfer torques
 that stem from uniform (${\bm v}_n$) and staggered (${\bm v}_\ell$) electron spin densities. 
 While the former is well recognized, the latter is often overlooked. 
 We show that both ${\bm v}_n$ and ${\bm v}_\ell$ contribute equally to the spin-wave Doppler shift.
 Microscopic calculations are presented for electrons on a two-dimensional square lattice 
with nearest-neighbor ($t$) and next-nearest-neighbor ($t'$) hopping, 
which interpolate two opposite transport regimes of strongly-coupled AF ($t'/t \ll 1$) 
and two weakly-coupled ferromagnets ($t'/t \gg 1$). 
 In the AF transport regime ($t'/t \ll 1$), ${\bm v}_n$ and ${\bm v}_\ell$ have opposite signs, 
and the sign of the Doppler shift depends on band filling; 
${\bm v}_n$ (${\bm v}_\ell$) is dominant near the AF gap (near the band bottom or the top).
 As $t'/t$ is increased, ${\bm v}_n$ undergoes a sign change  
whereas ${\bm v}_\ell$ does not. 
 In the limit of vanishing $t$, ${\bm v}_n$ and ${\bm v}_\ell$ coincide and the spin-transfer 
torque reduces to that of ferromagnets. 
}

\begin{document}
\maketitle

Spin waves are low-energy excitations of magnetically ordered systems 
that carry heat and angular momentum,
 and are expected to play important roles in spintronics. 
 Unlike electric currents that can also carry heat and angular momentum, 
 spin waves do not suffer Joule heating and can be more energy efficient. 
 Although much of the work done so far have focused on ferromagnets (FMs), 
 a new surge in the interest of antiferromagnetic (AF) spintronics sheds 
light on new possibilities of spin waves.

 Antiferromagnets (AFs) are a class of materials with many advantages over FMs, 
 such as the absence of leakage magnetic field, 
 rigidity to magnetic perturbations,
 fast spin dynamics, and diverse candidate materials \cite{MacDonald2011,Jungwirth2016,Gomonay2017,Baltz2018}. 
 The fast spin dynamics of AFs allows for 
 spin waves that can reach frequencies in the terahertz range,
 while ferromagnetic (FM) spin waves remain in the gigahertz range. 
 AF spin waves come with another  perk that might revolutionize magnonics,
 which is the chirality or isospin degree of freedom \cite{Daniels2018}. 
 In contrast to FM spin waves that can only encode information in the amplitude, 
 AF spin waves have multiple modes with different chiralities.
 Specifically, collinear AFs possess two degenerate eigenmodes with opposite chiralities. 
 Despite the highly anticipated features of AF spin waves, the means to control them remain limited.

 From a scalability perspective, electrical manipulation of spin waves is ideal. 
 In FMs, the effects of electric current on spin waves are relatively well known
  \cite{Lederer1966,Vlaminck2008,Seo2009,Sekiguchi2012,Chauleau2014}.  
 In particular, the (reactive) spin-transfer torque (STT) causes a Doppler shift on spin waves, 
 which can be used as a probe of spin-polarized transport in magnetic materials \cite{Vlaminck2008},  
 to realize effective black holes \cite{Roldan-Molina2017}, and so on. 
 Spin-wave Doppler shifts also realize in AFs \cite{Swaving2011}.
   Pioneering theories have elucidated key properties of AF spin torques phenomenologically
  \cite{Xu2008,Swaving2011,Hals2011,Tveten2013,Yamane2016,Park2020},
  and microscopic theories have identified their origins.\cite{Fujimoto2021,Nakane2021PRB}
 Experiments on AF spintronics remain limited as
 the lack of leakage magnetic field hinders detection of magnetic information,
  while the immunity to external fields forbids easy manipulation of the AF spins. 
  Compensated ferrimagnets have been used to overcome these difficulties, 
in which AF spin dynamics is realized at the angular momentum compensation temperature 
  while a macroscopic magnetic moment is finite allowing for detection and manipulation 
 of the magnetic texture \cite{Kim2017,Okuno2019,Kim2021}.

 In this Letter, we microscopically explore the effect of electric current on AF spin dynamics,
 and study the spin-wave properties in particular. 
 We first derive the equations of motion for AF spins 
 in terms of the N\'eel vector and the uniform magnetization, and identify 
  two kinds of  reactive STTs. 
      While only one STT acts on AF domain walls,\cite{Hals2011,Tveten2013,Nakane2021PRB} 
we found that the two STTs contribute to the spin-wave Doppler shift.
 This distinction is a new feature of AF STT, not present in FM. 
 The coefficients of the  two STTs are then calculated  based on a microscopic electron model. 
 With nearest-neighbor (n.n.) and next-nearest-neighbor (n.n.n.) electron hopping considered, 
the model incorporates two typical  transport regimes, 
namely, strongly-coupled AF and two decoupled FMs. 
 By interpolating these two limiting cases, 
 we find that the two torques are comparable in magnitude in general, 
 that they have opposite signs in the AF regime, and that they approach 
 a common expression (half of the FM STT) in the FM limit.

 We consider a metallic, two-sublattice AF consisting of localized spins and  
conduction electrons, interacting mutually via the s-d exchange interaction.   
 The Hamiltonian is 
\begin{align}
    H = H_S + H_{\rm el} + H_{\rm sd}  . 
\label{eq:H}
\end{align}
 The localized spins and their coupling to the electrons are described, respectively, by 
\begin{align}
    H_S &= J\sum_{\langle i,j \rangle}  {\bm S}_i\cdot{\bm S}_j  - K\sum_i (S_i^z)^2 , 
\\
    H_{\rm sd} &= 
    -J_{\rm sd} \sum_i {\bm S}_i \cdot c_i^\dagger {\bm\sigma} c_i  , 
\end{align}
where ${\bm S}_i$ is a classical spin at site $i$,
$J>0$ is the AF exchange coupling constant between the n.n. sites,
and $K>0$ is the easy-axis anisotropy constant.
 In $H_{\rm sd}$, $c_i^\dagger = (c_{i\uparrow}^\dagger,c_{i\downarrow}^\dagger)$ 
are electron creation operators at site $i$, $\bm\sigma$ is a vector of Pauli matrices, 
and $J_{\rm sd}$ is the s-d exchange coupling constant.
 The Hamiltonian $H_{\rm el}$ for the electrons will be specified later.

 We first describe the AF spin dynamics by considering a general bipartite lattice 
in $d$ spatial dimensions. 
 We adopt the exchange approximation, in which $J$ is considered the largest energy scale 
in the spin system.\cite{lifshitz_pitaevskii,Tveten2016}  
 Then the description is simplified in terms of the N\'eel vector, ${\bm n}$,  
and the uniform moment, ${\bm l}$.
 We introduce them by writing \cite{Haldane1983,com}  
\begin{align}
 {\bm S}_i = S \left\{ {\bm l}_i + (-)^i {\bm n}_i \right\} , 
\end{align}
where $S = |{\bm S}_i|$ is the constant magnitude 
and $(-)^i = \pm 1$ is the sublattice-dependent factor, 
and then by adopting the continuum approximation, 
${\bm n}_i \to {\bm n} ({\bm r})$ and ${\bm l}_i \to {\bm l} ({\bm r})$. 
 We assume spatial variations of ${\bm n}_i$ and ${\bm l}_i$ are slow throughout.  
 The apparent doubling of degrees of freedom is not harmful  
 if ${\bm n}$ and ${\bm l}$ are smooth enough and do not contain large-wavevector components.

The Lagrangian density for localized spins  in the continuum approximation is then given by
\begin{align}
&
  {\mathcal L}_S = 
s_n \big\{ {\bm l}\cdot({\bm n} \times \dot{\bm n}) 
              - {\mathcal H}_S - {\mathcal H}_{\rm sd} \big\} , 
\\
&\quad
    {\mathcal H}_S = 
    \frac{1}{2}
    \bigg\{ \tilde J \, {\bm l}^{\, 2} + \frac{c^2}{\tilde J} \sum_{i=1}^{d} (\partial_i{\bm n})^2 
             - \tilde K \, (n^z)^2 \bigg\} , 
\\&\quad
    {\mathcal H}_{\rm sd} = 
    -\frac{M}{s_n} ( {\bm l}\cdot\hat{\bm\sigma}_\ell + {\bm n}\cdot\hat{\bm\sigma}_n  )  , 
\end{align}
where $\hat{\bm\sigma}_n$ and $\hat{\bm\sigma}_\ell$ are the staggered and uniform spin densities of the conduction electrons. 
 We defined $\tilde{J} = 2zJS/\hbar$, $c = (zJSa)/(\hbar\sqrt{d})$, $\tilde{K} = 2SK/\hbar$, 
$s_n = 2\hbar S / (2a^d)$ and $M = J_{\rm sd} S$, 
where $z$ is the coordination number (number of n.n. sites)
  and $a$ is the lattice constant. 
 This leads to the equations of motion, 
\begin{align}
    \dot{\bm n} &=  {\bm H}_\ell \times{\bm n} + {\bm t}_n  , 
\label{eq:LLG_n}
\\
    \dot{{\bm l}} &= {\bm H}_n \times{\bm n} + {\bm H}_\ell \times{\bm l} + {\bm t}_\ell  , 
\label{eq:LLG_ell}
\end{align}
 with effective fields, ${\bm H}_n = \partial{\mathcal H}_S/\partial{\bm n}$ 
and ${\bm H}_\ell = \partial{\mathcal H}_S/\partial{\bm l}$, 
 and spin torques from the conduction electrons,  
\begin{align}
    {\bm t}_n &= \frac{M}{s_n} {\bm n}\times\langle\hat{\bm\sigma}_\ell\rangle  , 
\\
    {\bm t}_\ell &= \frac{M}{s_n}
    \big\{ \,   {\bm n} \times\langle\hat{\bm\sigma}_n\rangle
    + {\bm l}\times\langle\hat{\bm\sigma}_\ell \rangle  \, \big\} . 
\label{eq:t_ell}
\end{align}
 Note that Eqs. \eqref{eq:LLG_n} and \eqref{eq:LLG_ell} 
are consistent with the constraints, ${\bm l}\cdot {\bm n} = 0$ and $|{\bm n}|=1$.

 Under a current flow or time-dependent $\bm n$ and ${\bm l}$, 
the spin densities $\langle\hat{\bm\sigma}_n\rangle$ and $\langle\hat{\bm\sigma}_\ell\rangle$ 
are expected to acquire nonequilibrium components,   
\begin{align}
    \langle\hat{\bm\sigma}_n\rangle
&= \frac{s_n}{M} \big\{  
        - \beta_n ({\bm v}_n \!\cdot\! \nabla) \, {\bm n}
        + {\bm n} \times({\bm v}_\ell \!\cdot\! \nabla) \, {\bm l}
        -\alpha_n\dot{\bm n} \, 
    \big\}  , 
\label{eq:sigma_n}
\\
    \langle\hat{\bm\sigma}_\ell\rangle
&=  \frac{s_n}{M} \big\{ \, 
          {\bm n} \times({\bm v}_n \!\cdot\! \nabla) \, {\bm n}
        -\alpha_\ell  \dot {\bm l}  \, 
    \big\}  , 
\label{eq:sigma_ell}
\end{align}
where ${\bm v}_n$ and ${\bm v}_\ell$ are coefficients for the current-induced torques, 
$\beta_n$ characterizes the  dissipative STT (the so-called $\beta$-torque), 
and $\alpha_n$ and $\alpha_\ell$ are damping coefficients. 
 The microscopic expressions of ${\bm v}_n$, $\beta_n$, $\alpha_n$ and $\alpha_\ell$ 
have been derived in {Ref.~\citen{Nakane2021PRB}},
 but the ${\bm v}_\ell$-term was overlooked there \cite{com2}. 
 Here, we emphasize that the ${\bm v}_\ell$-term  
should be retained in the present (exchange) approximation. 
  In fact, Eq.~\eqref{eq:t_ell} shows that, in $\langle\hat{\bm\sigma}_n\rangle$, we need to retain terms 
which are one order higher in ${\bm l}$ compared to those in $\langle\hat{\bm\sigma}_\ell \rangle$. 
 Note that the above spin densities respect the sublattice symmetry 
${\bm n}\rightarrow -{\bm n}$, ${\bm l} \rightarrow {\bm l}$ 
in  Eqs. \eqref{eq:LLG_n} and \eqref{eq:LLG_ell}.
 Remaining terms that respect the sublattice symmetry,
 relevant in the exchange approximation,
 are ${\bm n}\times\dot{{\bm l}}$
 for  $ \langle\hat{\bm\sigma}_n\rangle $  
and ${\bm n}\times\dot{\bm n}$  
for  $  \langle\hat{\bm\sigma}_\ell\rangle $, 
 which only serve as renormalization of the coefficients on the left-hand side 
 of Eqs.~\eqref{eq:LLG_n} and \eqref{eq:LLG_ell}.

The equations of motion can then be written explicitly as
\begin{align}
    \dot{\bm n}
&=  \tilde{J}\,  {\bm l}\times{\bm n} -({\bm v}_n\cdot\nabla){\bm n}
\label{eq:LLG_n2}
\\
    \dot{{\bm l}}
&= - \big( \tilde{J}^{-1}c^2\nabla^2{\bm n} + \tilde{K}n^z\hat z  \big) \times {\bm n}
\nonumber \\
&\quad  +  \big\{ \beta_n ({\bm v}_n\cdot\nabla){\bm n} + \alpha_n \dot{\bm n} \big\}  \times{\bm n}
\nonumber \\
&\quad   - ({\bm v}_\ell\cdot\nabla) \, {\bm l}
    + {\bm n} [{\bm l} \cdot 
       ({\bm v}_n\cdot\nabla){\bm n} + {\bm n} \cdot ({\bm v}_\ell\cdot\nabla){\bm l}] . 
\label{eq:LLG_ell2}
\end{align}
 The terms with $\alpha_\ell$ are dropped in the exchange approximation. 
 One may eliminate ${\bm l}$ from these equations to obtain
\begin{align}
    {\bm n}\times\ddot{\bm n}
&=  {\bm n} \times \big( c^2\nabla^2{\bm n}  + \tilde{J} \tilde{K} n^z \hat z \big)
\nonumber \\
&\quad 
    - \tilde{J}  {\bm n} \times
    \big[   \beta_n ({\bm v}_n \cdot \nabla){\bm n} + \alpha_n \dot{\bm n} \big] 
\nonumber \\
&\quad  - {\bm n} \times [({\bm v}_n+{\bm v}_\ell)\cdot\nabla] \dot{\bm n}
\nonumber \\
&\quad  - {\bm n} \times ({\bm v}_\ell\cdot\nabla)({\bm v}_n\cdot\nabla) {\bm n}  . 
\end{align}
 Linearizing this equation around a uniform state ${\bm n} = \hat z$ 
by considering a small transverse component $\delta{\bm n}$ such that 
${\bm n}({\bm r}) = \hat{z} + \delta{\bm n} \, e^{i{\bm q}\cdot{\bm r}-i\omega t}$,  
one obtains a dispersion relation, 
\begin{align}
    \omega^2
&=   c^2 q^2  + \tilde{J}\tilde{K}
 + i \tilde{J} \, 
    \big\{ \beta_n ({\bm v}_n\cdot {\bm q})  - \omega\alpha_n  \big\}
\nonumber \\
&\quad  + [({\bm v}_n+{\bm v}_\ell)\cdot {\bm q}] \, \omega
    -({\bm v}_\ell\cdot{\bm q})({\bm v}_n\cdot{\bm q})  .
\end{align}
 Solving for $\omega$ to the leading order in ${\bm v}_n$ and ${\bm v}_\ell$  gives   
\begin{align}
  \omega
&=  \bigg( c^2 q^2  + \tilde{J}\tilde{K} + \tilde{J}\beta_n ({\bm v}_n\cdot i{\bm q})
        -({\bm v}_\ell\cdot{\bm q})({\bm v}_n\cdot{\bm q})
\nonumber \\
&\qquad\quad
        + \frac{[ i \alpha_n \tilde{J} - ({\bm v}_n + {\bm v}_\ell) \cdot {\bm q}]^2}{4}
    \bigg)^{1/2}
\nonumber \\
&\quad  \mp\frac{i\alpha_n\tilde{J}-({\bm v}_n+{\bm v}_\ell)\cdot {\bm q}}{2}
\\ 
&\simeq   \sqrt{ c^2 q^2 + \tilde{J}\tilde{K}  }
    \pm \frac{({\bm v}_n + {\bm v}_\ell)\cdot {\bm q}}{2}  . 
\label{eq:SW_dispersion}
\end{align}
 In the last expression, we dropped the effects of damping and dissipative $\beta$-torques.
 We see that the Doppler shift of AF spin waves is given by
$({\bm v}_n+{\bm v}_\ell)/2$.  
 Thus the Doppler shift is caused by the two torques of different origin, 
the ${\bm v}_n$- and the ${\bm v}_\ell$-terms. 
 Because they contribute to the Doppler shift, we identify both of these torques to be STTs in AFs.
This is one of the main results of this Letter.
  A similar Doppler shift (with a factor of 1/2) was obtained in Ref.~\citen{Swaving2011} 
but without account for ${\bm v}_\ell$.

 To determine the magnitude of ${\bm v}_n$ and ${\bm v}_\ell$, 
we next perform a microscopic calculation. 
 To be explicit, we consider electrons on a two-dimensional square lattice, 
\begin{align}
    {H}_{\rm el}
&=  -t \sum_{\langle i,j \rangle}  (c_i^\dagger c_j + {\rm H.c.})
    - t' \sum_{\langle\!\langle l,m \rangle\!\rangle}  (c_l^\dagger c_m + {\rm H.c.})
    + V_{\rm imp}  , 
\label{eq:H_el}
\end{align}
with n.n. hopping (first term), n.n.n. hopping (second term), and subject to impurity potentials 
(last term). 
 To calculate ${\bm v}_n$ and ${\bm v}_\ell$, it is sufficient to consider nonmagnetic impurities,  
 $V_{\rm imp} = u_{\rm i} \sum'_j  c_j^\dagger c_j$,
where $u_{\rm i}$ is the strength of the impurity potential and 
 the sum is taken over the impurity positions. 
 The number of impurities is assumed equal for the two sublattices, with density $n_{\rm i}$.
 Combining $H_{\rm el}$ with $H_{\rm sd}$ completes the model for the conduction electrons. 

 To treat the spatial variation of the N\'eel vector, we employ the method of spin gauge field \cite{Tatara2008}. 
 We perform a local ${\rm SU}(2)$ rotation $U_i$ that brings the N\'eel vector at each site $i$ to the $z$ direction, 
$U_i^\dagger ({\bm n}_i \cdot{\bm\sigma}) U_i = \sigma^z$.
 The hopping term is then modified through $ U_i^\dagger U_j = e^{iA_{ij}}$, 
which introduces the spin gauge field $A_{ij}$. 
 We also define the corresponding $3\times 3$ rotation matrix ${\mathcal R}_i$
by $ U_i^\dagger {\bm\sigma} U_i = {\mathcal R}_i {\bm\sigma}$.

 The Hamiltonian then becomes   
$H_{\rm el} + H_{\rm sd} =  H_0 + H' + V_{\rm imp}$, 
\begin{align}
    H_0 &=  \sum_{ i,j }  t_{ij} \tilde{c}_i^\dagger \tilde{c}_j  
    - M  \sum_i (-)^i \tilde{c}_i^\dagger \sigma^z \tilde{c}_i  ,
\\
    H' 
&= \sum_{ i,j }  t_{ij} \tilde{c}_i^\dagger i A_{ij} \tilde{c}_j  
    -M \sum_i ( {\mathcal R}_i^{-1} {\bm l}_i )\cdot \tilde{c}_i^\dagger {\bm\sigma} \tilde{c}_i  , 
\label{eq:H'}
\end{align}
up to $O(A_{ij})$,   
where $\tilde{c}_i = U_i^\dagger c_i$ is defined in the rotated frame, 
and $t_{ij}$ is $-t$ ($-t'$) if $i,j$ are n.n. (n.n.n.) pairs and zero otherwise. 
 Since $A_{ij}$ and ${\bm l}$ are considered small, we treat $H'$ perturbatively.\cite{Nakane2020}

 The unperturbed part $H_0$ describes electrons under a uniform AF moment, 
  whose dispersion is 
\begin{align}
 E_{{\bm k}, \pm} =  \pm E_{\bm k} + \varepsilon'_{\bm k} , 
\end{align}
where $E_{\bm k} = \sqrt{\varepsilon_{\bm k}^2 + M^2}$, 
$\varepsilon_{\bm k} = -2t (\cos k_x + \cos k_y)$ comes from the n.n. hopping, 
and $\varepsilon'_{\bm k} = -4t' \cos k_x \cos k_y$ from the n.n.n. hopping. 
 At $t'=0$, it reduces to $E_{{\bm k}, \pm} =  \pm E_{\bm k}$. 
 For $t=0$, it becomes $E_{{\bm k}, \pm} =  \varepsilon'_{\bm k} \pm |M|$, 
and the model describes two decoupled FMs with opposite magnetization  
(hence vanishing total magnetization). 
 We call the former the \lq\lq AF transport limit'', and the latter the \lq\lq FM transport limit''. 
 More generally, the model is in the \lq\lq AF transport regime'' for $t \gg t'$, 
and in the \lq\lq FM transport regime'' for $t \ll t'$.

  In the following, we calculate the electron spin densities, 
$\langle\hat{\bm\sigma}_n\rangle$ and $\langle\hat{\bm\sigma}_\ell\rangle$, 
in response to an applied electric field ${\bm E}$,  
using the linear response theory and the Green's function method,  
\cite{Kohno2006,Kohno2007,Nakane2021PRB}
  and identify the coefficients, 
 ${\bm v}_n$ in Eq.~(\ref{eq:sigma_ell}) and ${\bm v}_\ell$ in Eq.~(\ref{eq:sigma_n}). 
 The effects of impurities are considered in the Born approximation 
together with ladder vertex corrections. 
 Details of the calculation are presented in the Supplemental Material (SM) \cite{SM}.

 The STT parameter ${\bm v}_n$ that arises through the uniform spin density 
$\langle\hat{\bm\sigma}_\ell\rangle$ is obtained as 
\begin{align}
    {\bm v}_n &= 
    \frac{ -e\hbar {\bm E}}{s_n} 
    M^2
    \Lambda
    \sum_{\eta = \pm 1}\frac{1}{N}\sum_{\bm k}
    \frac{(v_i^0)^2 - (v_i')^2}{2E_{\bm k} |\eta E_{\bm k}\gamma_0 + M\gamma_3|}
\nonumber \\&\qquad
    \times
     \frac{\eta E_{\bm k}\gamma_3 + M\gamma_0}{\eta E_{\bm k}\gamma_0 + M\gamma_3}
     \delta(\mu-\varepsilon'_{\bm k}-\eta E_{\bm k})  , 
\label{eq:vn_result}
\end{align}
where $v^0_i = \partial_i \varepsilon_{\bm k}$ is the velocity coming from the n.n. hopping, 
and $v'_i = \partial_i \varepsilon'_{\bm k}$ from the n.n.n. hopping. 
 The effects of damping  surface through 
$\gamma_0=\pi n_{\rm i} u_{\rm i}^2\sum_{\eta=\pm 1}\frac{1}{N}\sum_{\bm k}\delta(\mu-\varepsilon'_{\bm k}-\eta E_{\bm k})$, 
$\gamma_3=\pi n_{\rm i} u_{\rm i}^2 M\sum_{\eta=\pm 1}\frac{1}{N}\sum_{\bm k}
\delta(\mu-\varepsilon'_{\bm k}-\eta E_{\bm k})/({\eta E_{\bm k}})$,
and
the impurity ladder vertex correction 
$\Lambda = (1-\Lambda_1)^{-1}$, where 
\begin{align}
 \Lambda_1 &= 
    \frac{\pi n_{\rm i}u_{\rm i}^2}{\gamma_0}
    \sum_{\eta = \pm 1}  \frac{1}{N}\sum_{\bm k}
    \frac{\varepsilon_{\bm k}^2}{E_{\bm k}}
    \frac{\delta(\mu-\varepsilon'_{\bm k}-\eta E_{\bm k})}{|\eta E_{\bm k} + M(\gamma_3/\gamma_0)|}
,
\end{align}
$N$ is the total number of sites, 
and the chemical potential $\mu$ is measured from the AF gap center at $t'=0$.

\begin{figure}[tbp]
    \centering
    \includegraphics[width=0.47\textwidth]{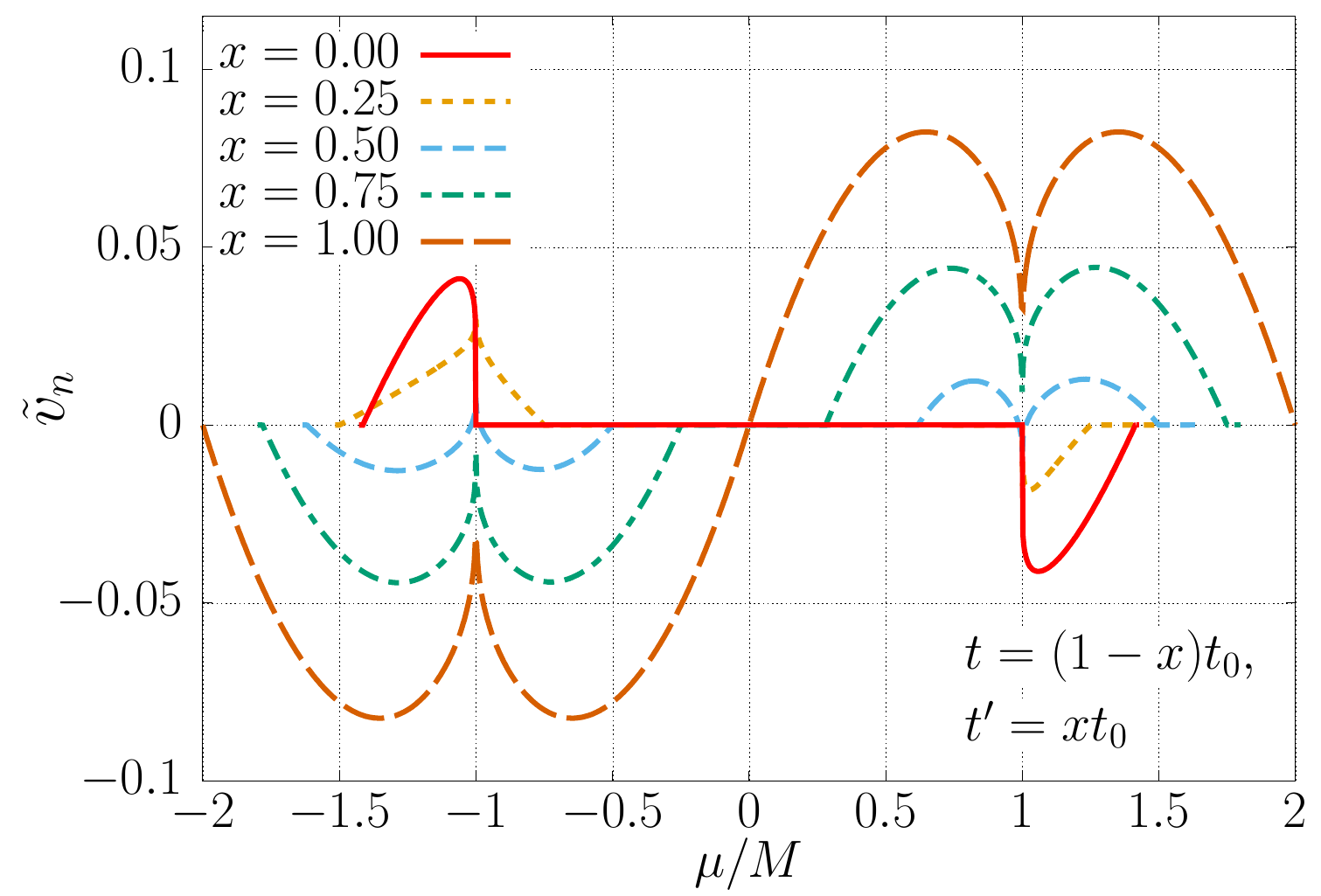}
    \includegraphics[width=0.47\textwidth]{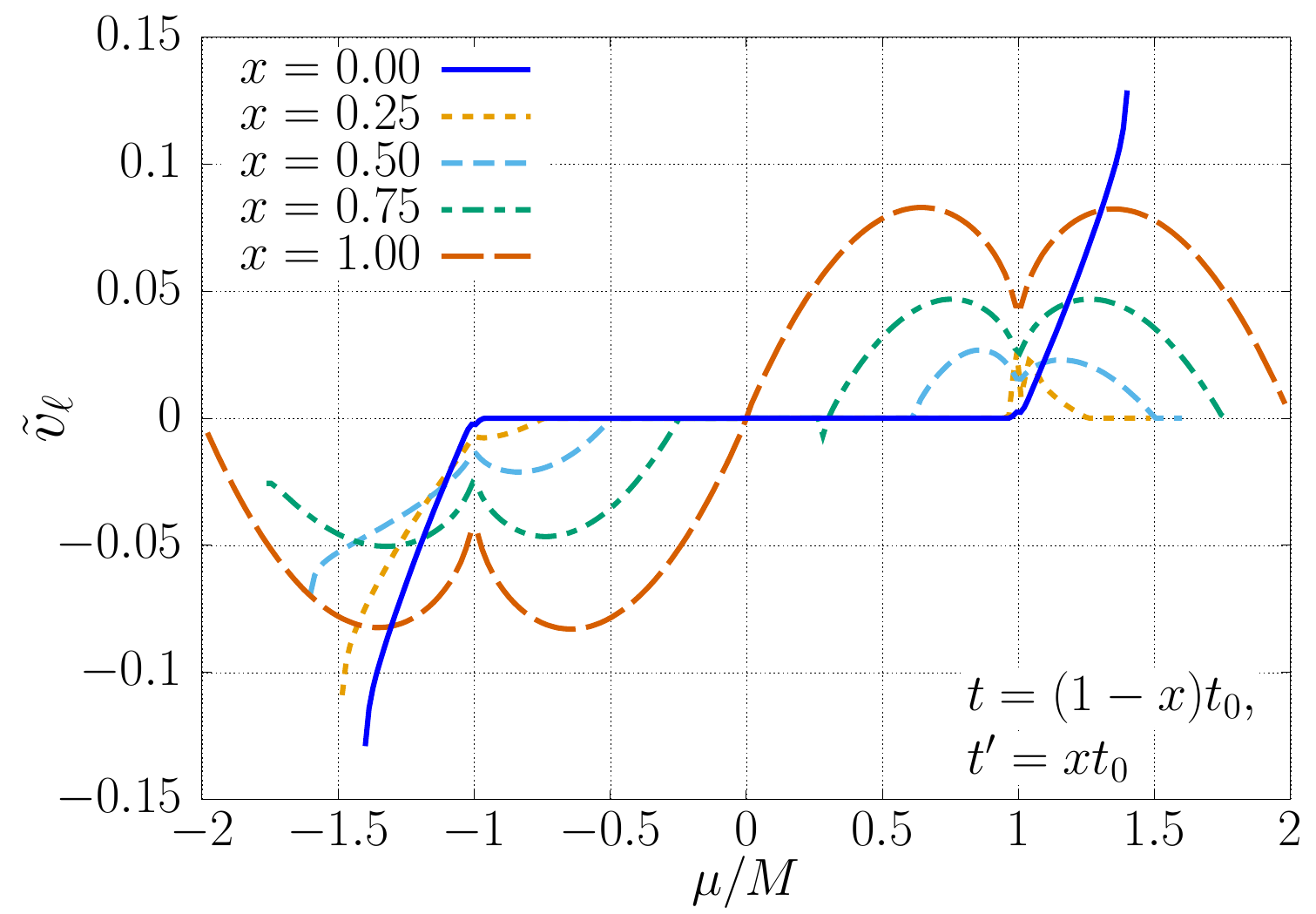}
    \caption{(Color online) The STT coefficients, $v_n$ (upper panel) and $v_\ell$ (lower panel), 
    as functions of chemical potential $\mu$ for several choices of $x$, 
    where $t = (1-x) \, t_0 $ and $t' = x \, t_0$ with $t_0 / M = 0.25$. 
     Plotted are the normalized values, 
    $\tilde{v}_n = ({\bm v}_n \cdot {\bm E}) s_n \tilde{\gamma} / (e\hbar |{\bm E}|^2)$ and 
    $\tilde{v}_\ell = ({\bm v}_\ell \cdot {\bm E}) s_n \tilde{\gamma} / (e\hbar |{\bm E}|^2)$, 
   where 
    $\tilde{\gamma} = \pi n_{\rm i} u_{\rm i}^2 / M^2$ is the dimensionless damping parameter. 
    The choice $t_0 / M = 0.25$ corresponds to a \lq\lq strong AF'', in which 
   the upper and lower bands do not overlap.
    The following features are seen. 
     (i) $v_n$ changes sign as $x$ is increased from $x=0$ (AF transport limit) 
    to $x=1$ (FM transport limit), whereas $v_\ell$ keeps the same sign throughout. 
    (ii) $v_\ell$ coincides with $v_n$ at $x=1$.
    (iii) $v_n$ and $v_\ell$ are odd functions of $\mu$ at $x=0$ and $1$
    because of the presence of particle-hole symmetry, but not for general $x$. 
    }
\label{fig:fig1}
\end{figure}

 Equation (\ref{eq:vn_result}) holds for arbitrary values of $t$ and $t'$, 
and incorporate the two opposite transport regimes of AF and FM. 
 In the AF transport limit, $t' = 0$, one has $\Lambda = (\mu^2+M^2)/(2M^2)$, 
and ${\bm v}_n$ reduces to 
\begin{align}
    {\bm v}_n 
=  -\frac{\hbar}{2es_n}\, 
    \frac{\mu M}{\mu^2-M^2}\, \sigma_{xx}{\bm E}  \qquad (t' = 0)  , 
\label{eq:AF_stt}
\end{align}
where $\sigma_{xx} = 2e^2 D \nu$ is the longitudinal conductivity, 
$D = \frac{1}{N}\sum_{\bm k} (\partial_x E_{\bm k})^2 \delta(|\mu| - E_{\bm k}) \tau/ \nu $ 
 is the diffusion constant, 
$\tau = (1/2) (\gamma_0+M\gamma_3/\mu )^{-1}$ is the scattering time,  
and 
$\nu = \frac{1}{N} \sum_{\bm k} \delta(|\mu| - E_{\bm k})$ is the density of states per spin, 
all evaluated at $t' = 0$. 
This result agrees with the one reported in Ref.~\citen{Nakane2021PRB}.

 In the opposite limit, $t=0$, we retrieve the STT for FMs \cite{Kohno2006,Kohno2007}
\begin{align}
    {\bm v}_n 
=  -\frac{\hbar}{2es_n} 
    \,  (\sigma_\uparrow-\sigma_\downarrow) \, {\bm E}  \qquad (t = 0)  , 
\label{eq:FM_stt}
\end{align}
where  $\sigma_{\uparrow}$ ($\sigma_{\downarrow}$) 
is the longitudinal conductivity 
of electrons in band  $\varepsilon'_{\bm k} - M$ ($\varepsilon'_{\bm k} + M$).
 We find that the sign of the STT in the AF transport regime, Eq. \eqref{eq:AF_stt}, 
is opposite to that in the FM transport regime, Eq. \eqref{eq:FM_stt}.

 To see how ${\bm v}_n$ develops between the AF and the FM regimes, 
we evaluate Eq.~(\ref{eq:vn_result}) numerically, and plot the result 
  in the upper panel in Fig.~\ref{fig:fig1}. 
 We set $t = (1-x) \, t_0$ and $t' = x \, t_0$, 
which interpolate the AF transport regime ($x \sim 0$) 
and the FM transport regime ($x \sim 1$). 
 As seen, $v_n$ changes sign as $x$ is increased from $x=0$ to 1. 
 This means that the STT due to ${\bm v}_n$ has opposite sign between FM and AF. 
 This fact has been used in Ref.~\citen{Nakane2021PRB} to interpret the experimental result  
of domain wall motion in a compensated ferrimagnet GdFeCo, \cite{Okuno2019} 
which is expected to be in the AF transport regime \cite{Park2021}.

\begin{figure}[tbp]
    \centering
    \includegraphics[width=0.47\textwidth]{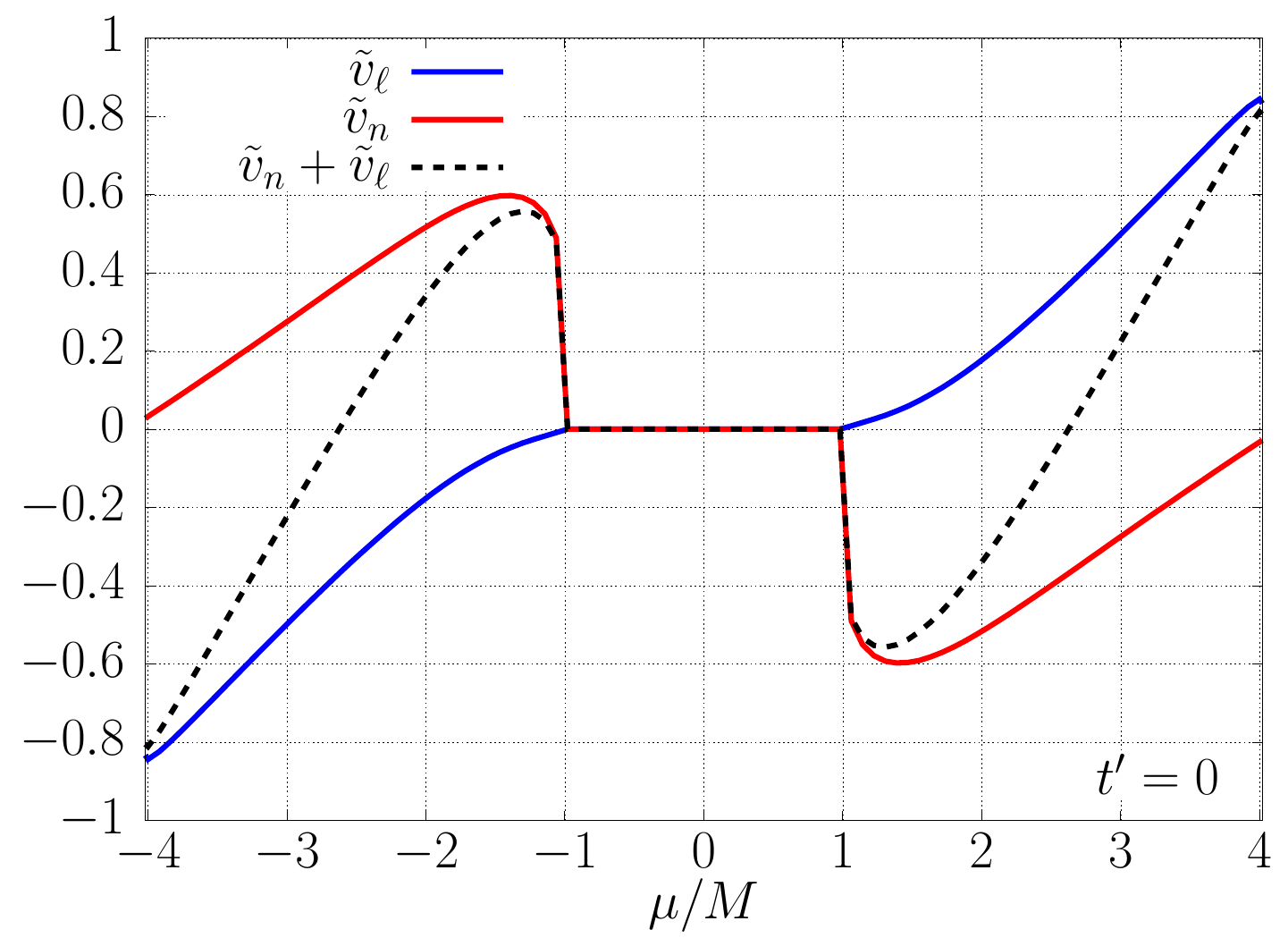}
    \caption{(Color online) Normalized STT coefficients, 
                 $\tilde{v}_n$ (red), $\tilde{v}_\ell$ (blue) and $\tilde{v}_n + \tilde{v}_\ell$ (black, dashed), 
              calculated with $t / M = 1$ and $t' = 0$.
    The total Doppler shift $v_n + v_\ell$ changes sign as a function of $\mu$. 
    }
\label{fig:fig2}
\end{figure}

 The STT ${\bm v}_\ell$ arising from the staggered spin density 
$\langle \hat {\bm\sigma}_n \rangle$
is calculated by considering the  perturbation by the 
 canting moment ${\bm l}$ \cite{Nakane2020}. 
Unlike the terms with $\beta_n$ and $\alpha_n$ in Eq. \eqref{eq:sigma_n}
 that arise in the presence of spin relaxation, 
 the ${\bm v}_\ell$-term does not require spin relaxation because it is  a reactive STT.
 In the AF transport limit $t'=0$, we obtain 
\begin{align}
    {\bm v}_\ell
&=  \frac{ \hbar}{2es_n}  \,  M\tau  \,
    \bigg[   \frac{\mu^2-M^2}{\mu D d} + 2 \zeta  \bigg]  \, \sigma_{xx}  {\bm E}  \ \ \ \ \  (t'=0), 
\end{align}
to leading order in $\tau$,
where $\zeta$ comes from the impurity correction to the ${\bm l}$-vertex 
(see Fig.~S1 (c) and Eq.~(S12) in SM\cite{SM}). 
 Like ${\bm v}_n$, ${\bm v}_\ell$ is an odd function of $\mu$
 because of opposite spin directions between the upper and lower electron bands, 
 and an odd function of $M$ for a similar reason.
 As seen  in Fig. \ref{fig:fig1} (blue line in the lower panel) and Fig. \ref{fig:fig2}, 
the sign of ${\bm v}_\ell$ relative to ${\bm E}$ is negative 
(hence positive relative to the electron flow) in the lower band, 
and ${\bm v}_\ell$ remains finite at the band bottom. 
 These are in contrast to ${\bm v}_n$. 
 As a result,  near the band bottom,  the total Doppler shift $({\bm v}_n+{\bm v}_\ell)/2$ 
is dominated by ${\bm v}_\ell$   and  hence  negative.  
 As the chemical potential is shifted to the AF gap edge, 
${\bm v}_n$ starts to take over and the Doppler shift undergoes a sign change.

 In the FM transport limit $t=0$, we find 
\begin{align}
    {\bm v}_\ell
=  -\frac{\hbar}{2es_n} (\sigma_\uparrow-\sigma_\downarrow) \, {\bm E}   \qquad (t = 0) , 
\label{eq:vl_FM}
\end{align}
which coincides with ${\bm v}_{n}$ in Eq. \eqref{eq:FM_stt}. 
This fact also indicates that ${\bm v}_\ell$ needs to be retained as the second STT in AFs.
 Thus, the AF spin waves receive a Doppler shift by 
$ ({\bm v}_n+{\bm v}_\ell)/2 = {\bm v}_\ell $, 
and this is exactly the Doppler shift in FM.

 For other $t$, $t'$ values, we have numerically evaluated Eqs.~(S27)-(S30) given in SM \cite{SM}, 
and the results are plotted in the lower panel in Fig. \ref{fig:fig1}. 
 In contrast to ${\bm v}_n$, it does not change sign with $x$, hence its sign is always that of FM.

 While both ${\bm v}_n$ and ${\bm v}_\ell$ appear in the spin-wave Doppler shift, 
only ${\bm v}_n$ appears in the collective-coordinate equations of AF domain wall motion. 
\cite{Hals2011,Tveten2013,Nakane2021PRB} 
 Thus, unlike FM in which there is only one kind of STT, 
AFs allow for two kinds of STTs that play different roles depending on 
 physical phenomena.

 In this Letter, we have studied STTs in AF that induce a Doppler shift in spin-wave spectrum. 
  We have shown that the Doppler shift in AFs is induced by two kinds of  reactive STTs, 
identified as ${\bm v}_n$ and ${\bm v}_\ell$, 
which arise through uniform and staggered electron spin densities, respectively, 
and are proportional to the spatial gradient of the N\'eel vector and the uniform moment, respectively. 
 Both STTs contribute to the spin-wave Doppler shift equally. 
 This contrasts with the effects on AF domain walls, to which only ${\bm v}_n$ is relevant.

 We next determined the STTs microscopically using a tight-binding model  
with n.n. and n.n.n. hopping. 
In the AF transport regime dominated by n.n. hopping, 
${\bm v}_n$ and ${\bm v}_\ell$ have opposite signs, and the sign of the Doppler shift depends on band filling.  
 Especially, ${\bm v}_\ell$ dominates near the band bottom (top), 
and the sign of the Doppler shift is negative (positive) relative to the applied field ${\bm E}$. 
 As the chemical potential is moved toward the AF gap, 
${\bm v}_n$ starts to dominate and the Doppler shift changes sign.
 In the FM transport limit with only the n.n.n. hopping,
both ${\bm v}_n$ and ${\bm v}_\ell$ coincide with the well-known STT in FM, 
and add up to reproduce the Doppler shift in FM.

\begin{acknowledgement}
 This work was partly supported by JSPS KAKENHI Grant Numbers JP15H05702, JP17H02929 and JP19K03744, 
and the Center of Spintronics Research Network of Japan. 
 JJN is supported by a Program for Leading Graduate Schools ``Integrative Graduate Education and Research in Green Natural Sciences'' 
 and Grant-in-Aid for JSPS Research Fellow Grant Number 19J23587. 
\end{acknowledgement}

\bibliographystyle{jpsj}
\bibliography{ref}

\end{document}


\renewcommand{\theequation}{S\arabic{equation}}

\maketitle

 In this Supplemental Material, we present the calculation of the uniform spin density 
${\mathcal R}^{-1}\langle\hat{\bm \sigma}_\ell\rangle$ and the staggered spin density 
${\mathcal R}^{-1}\langle\hat{\bm\sigma}_n\rangle$ (in the rotated frame), or 
$\langle\hat{\bm \sigma}_\ell\rangle$ and 
$\langle\hat{\bm\sigma}_n\rangle$ (in the original frame), 
in response to an external electric field, to obtain ${\bm v}_n$ and ${\bm v}_l$, 
respectively.

\section{Green's function} 

 The unperturbed Hamiltonian is written as
\begin{align}
    H_0 =  \sum_{\bm k}  c_{\bm k}^\dagger h_{\bm k} c_{\bm k}  , 
\end{align}
where $c_{\bm k} = {}^t 
(c_{\bm k\uparrow, A},c_{\bm k\downarrow, A},c_{\bm k\uparrow, B},c_{\bm k\downarrow, B})$
and 
\begin{align}
    h_{\bm k}
&=  \varepsilon'_{\bm k}  + \varepsilon_{\bm k}\tau_1 - M\sigma^z\tau_z . 
\end{align}
 The Pauli matrices that act in the sublattice (spin) space are denoted 
by ${\bm \tau}$ (${\bm \sigma}$).
 The retarded Green's function of the unperturbed Hamiltonian is 
\begin{align}
    G^R = \mu^R + T^R\tau_1 + J^R\sigma^3\tau_3 ,
\end{align}
where 
$\mu^R = (\mu-\varepsilon'_{\bm k}+i\gamma_0)/D^R$, 
$T^R = \varepsilon_{\bm k}/D^R$, 
$J^R = (-M+i\gamma_3)/D^R$, and
$D^R = (\mu-\varepsilon'_{\bm k}+i\gamma_0)^2-\varepsilon_{\bm k}^2 - (-M+i\gamma_3)^2$, 
and the advanced Green's function is given by its Hermitian conjugate, $G^A = (G^R)^\dagger$.
The velocity operator is given by  
\begin{align}
  v_i = v^0_i \tau_1 + v'_i  . 
\end{align}
 The spin gauge field appears through the perturbative Hamiltonian, 
$H_A = \sum_{\bm k}c_{\bm k}^\dagger v_i {A}_i c_{\bm k}$, and the anomalous velocity,
$ (\partial_j v_i) A_j$.
The spin gauge field is expanded using Pauli matrices as $A_i = {\bm A}_i\cdot{\bm\sigma}/2$,
where the perpendicular component of the spin gauge field is given by
${\bm A}_i-A^z_i\hat{z}=- {\mathcal R}^{-1} ({\bm n}\times\partial_i{\bm n})$.
 
The effect of $V_{\rm imp}$ is considered in the Born approximation and ladder type vertex 
correction, as shown in Fig. \ref{fig:fig3} (a), (b), and (c).

\section{Calculation of ${\bm v}_n$} 

 Here, we calculate the uniform spin density ${\mathcal R}^{-1}\langle\hat{\bm \sigma}_\ell\rangle$. 
  The first two diagrams in Fig. \ref{fig:fig3} are given by
\begin{align}
    {\rm (e1)+(e2)}
&=
    \frac{1}{2\pi} \frac{1}{N}  \sum_{\bm k}
    {\rm tr}  \big[  {\sigma}^\alpha  G^R  v_i A_i G^R  v_i(-e E_i)  G^A \big] + {\rm c.c.} , 
\end{align}
and the third diagram in Fig.  \ref{fig:fig3} is given by
\begin{align}
    {\rm (e3)}
&=
    \frac{1}{2\pi} \frac{1}{N}  \sum_{\bm k}
    {\rm tr}  \big[  {\sigma}^\alpha G^R  (\partial_iv_i)A_i(-e E_i)  G^A \big]  .
\end{align}
 Adding up the terms (e1), (e2), and (e3) determines the spin density $({\mathcal R}^{-1}{\bm\sigma}_\ell)^\alpha$ 
in response to an electric field $E_i$, 
without the effect of vertex correction taken into account.
Note that we are only interested in $\alpha = x,y$ because 
$\hat{\bm\sigma}_\ell\perp {\bm n}$.
From this it follows that the $z$ component of the spin gauge field $A_i^z$
does not contribute.

\renewcommand{\thefigure}{S\arabic{figure}}
\setcounter{figure}{0}
\begin{figure*}[tb]
    \centering
    \includegraphics[width=0.95\textwidth]{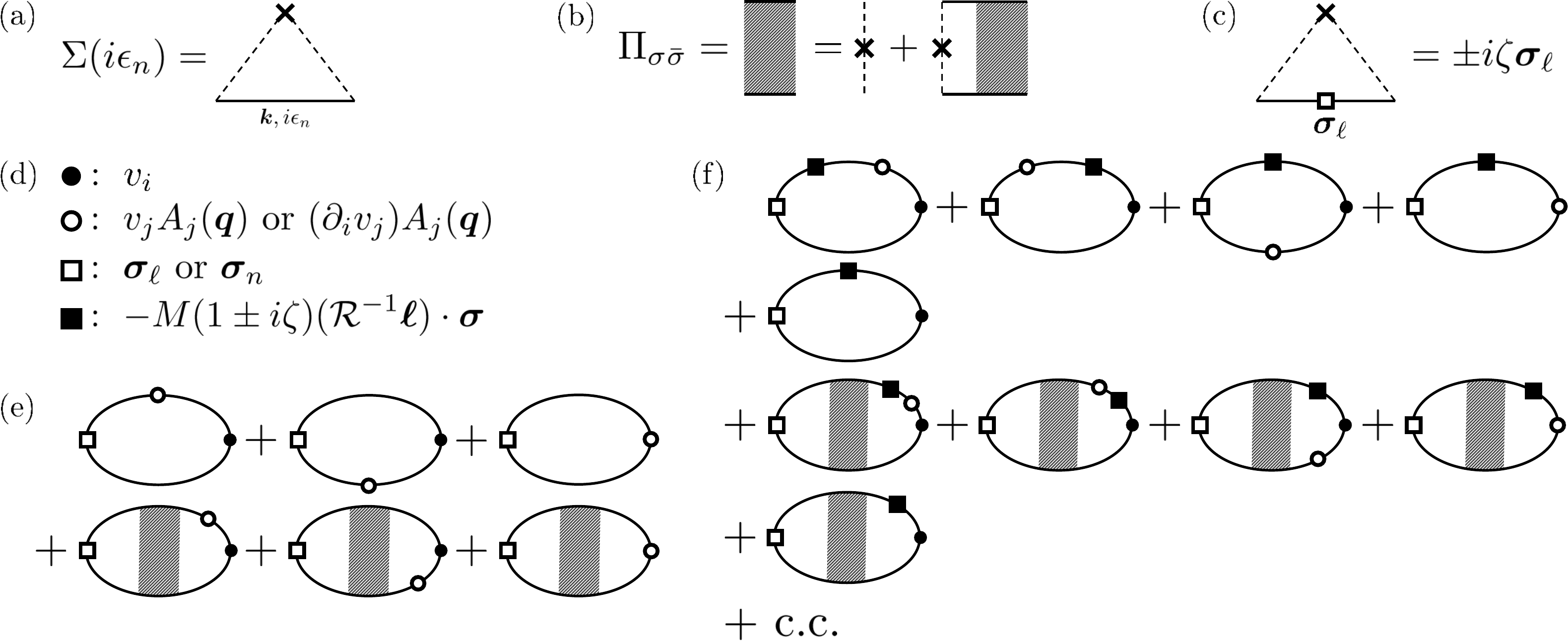}
    \caption{
     Feynman diagrams considered in the calculation. 
     The impurities are treated as in (a)-(c), and the calculated electron spin polarizations 
     are shown in (e) and (f), with the vertices defined in (d). 
     The solid line represents the electron Green's function, 
     and the dashed line with a cross represents nonmagnetic impurity scattering.  
        (a) Self-energy in the Born approximation. 
        (b) Four-point vertex in the ladder approximation. 
        The upper (lower) electron line is in the retarded (advanced) branch and has spin $\sigma$ ($\bar\sigma$).
        (c) Impurity correction to the (static) ${\bm l}$-vertex. 
            The plus (minus) sign is for the retarded (advanced) branch.
     (d) Definition of vertices. 
    The filled circle ($\bullet$) represents the current vertex $v_i^0$ that couples to the external electric field $E_i$. 
    The empty circle ($\circ$) is associated with the spin gauge field ${A}_i$,  
     coming either from the perturbation Hamiltonian, $v_l {A}_l ({\bm q})$, 
    or from the current vertex,  $(\partial_i v_i ) {\bm A}_i ({\bm q})$. 
    The empty square ($\square$) represents the uniform (${\bm \sigma}_{\ell}$) 
     or staggered (${\bm \sigma}_{n}$) spin density. 
    The filled square ($\blacksquare$) is the ${\bm l}$-vertex with impurity correction. 
    In (e) and (f), the right vertex represents the charge current that couples to ${\bm E}$, 
    and the upper (lower) electron lines are in the retarded (advanced) branch. 
        (e) Diagrams for the \lq\lq uniform'' spin density, ${\bm \sigma}_{\ell}$,  
            in first order in $A_i$.
        (f) Diagrams for the \lq\lq staggered'' spin density, ${\bm \sigma}_{n}$, 
          in first order in ${\bm l}$. 
        The diagrams with $\circ$ (spin gauge field) only give gauge-noninvariant terms ($\propto A^z_i$), 
       which cancel those arising from $(\partial_i {\mathcal R}^{-1}) \, {\bm l}$ in the remaining diagrams.
    }
\label{fig:fig3}
\end{figure*}

Integrating by parts, and using the relation $\partial_i G^R = G^R v_i G^R$,
(e3) can be merged with (e1)+(e2), 
\begin{align}
    {\rm (e1)+(e2)+(e3)}
&=
    \frac{1}{4\pi} A^\beta_i  \frac{1}{N}  \sum_{\bm k}
    \Big[
    (v_i^0)^2 \,  {\rm tr}
    \big\{  {\sigma}^\alpha  G^R  \tau_1 ( \sigma^\beta G^R  - G^R  \sigma^\beta  )  \tau_1  G^A \big\}
\nonumber \\
&\qquad\qquad\qquad
    +(v'_i)^2 \, 
    {\rm tr}
    \big\{  {\sigma}^\alpha  G^R ( \sigma^\beta G^R - G^R \sigma^\beta ) G^A  \big\}
    \Big](-e E_i) 
 + {\rm c.c.}
\end{align}
 Evaluating the trace and approximating the product of the Green's functions 
by a delta function, one obtains
\begin{align}
&
    {\rm (e1)+(e2)+(e3)}
\nonumber \\
&=
    -MA_i^\alpha
    \sum_{\eta = \pm 1}\frac{1}{N}\sum_{\bm k}
    \frac{(v_i^0)^2 - (v_i')^2}{2E_{\bm k} |\eta E_{\bm k}\gamma_0 + M\gamma_3|}
     \frac{\eta E_{\bm k}\gamma_3 + M\gamma_0}{\eta E_{\bm k}\gamma_0 + M\gamma_3}
     \delta(\mu-\varepsilon'_{\bm k}-\eta E_{\bm k})
    (-eE_i)  . 
\end{align}
 These are for diagrams without vertex corrections. 
 The effect of vertex corrections is taken into account by multiplying the above result by $\Lambda$, 
and we can now write
 the current induced uniform spin density as
\begin{align}
    \langle {\bm\sigma}_\ell \rangle
&=
    M{\bm n}\times\partial_i {\bm n}
    \,
    \Lambda
    \sum_{\eta = \pm 1}\frac{1}{N}\sum_{\bm k}
    \frac{(v_i^0)^2 - (v_i')^2}{2E_{\bm k} |\eta E_{\bm k}\gamma_0 + M\gamma_3|}
     \frac{\eta E_{\bm k}\gamma_3 + M\gamma_0}{\eta E_{\bm k}\gamma_0 + M\gamma_3}
     \delta(\mu-\varepsilon'_{\bm k}-\eta E_{\bm k})
    (-eE_i)  . 
\end{align}

\section{Calculation of ${\bm v}_\ell$}

 Here, we calculate the staggered spin density 
${\mathcal R}^{-1}\langle\hat{\bm\sigma}_n\rangle$
in response to an external electric field, and obtain ${\bm v}_\ell$. 
 We are interested in the first-order terms in the uniform moment ${\bm l}$, 
perturbed through 
\begin{align}
    H_\ell = -M \sum_{\bm k} ({\mathcal R}^{-1}{\bm l})\cdot{\bm\sigma}  . 
\end{align}
 The relevant diagrams are depicted in Fig. \ref{fig:fig3} (f). 
 The first four diagrams (without vertex correction) give 
\begin{align}
&  {\rm (f1) + \cdots + (f4)}
\nonumber \\
& = \frac{1}{2\pi} 
    (1+i\zeta)(-M{\mathcal R}^{-1}  {\bm l} \, )^\beta (-eE_i) 
\nonumber \\
& \ \ \times     \frac{1}{N}
    \sum_{\bm k}
    \Big\{ {\rm tr}
    \big[
        {\bm\sigma}^\perp\tau_3 G^R \sigma^\beta G^R v_j A_j  G^R v_i G^A
    \big]
 +
    {\rm tr}
    \big[
        {\bm\sigma}^\perp\tau_3 G^R v_j A_j  G^R \sigma^\beta G^R v_i  G^A
    \big]
\nonumber \\
&\qquad\qquad +
    {\rm tr}
    \big[
        {\bm\sigma}^\perp\tau_3 G^R  \sigma^\beta G^R  v_i  G^A v_j A_j G^A
    \big]
 +
    {\rm tr}
    \big[
        {\bm\sigma}^\perp\tau_3 G^R \sigma^\beta G^R  (\partial_j v_i) A_j G^A
    \big]
    \Big\}  , 
\end{align}
where $\zeta$ comes from the impurity correction to the ${\bm l}$-vertex, Fig. \ref{fig:fig3} (c), 
\begin{align}
    \zeta =  n_{\rm i} u_{\rm i}^2 \,
    {\rm Im} \bigg[ \frac{2}{N} \sum_{\bm k}   (\mu^R)^2 + (T^R)^2 -(J^R)^2 \bigg] . 
\label{eq:zeta}
\end{align}
 The trace is nonvanishing only for $A^z \sigma^z/2$ in $A_j$, which is, however, not gauge-invariant. 
 Note also that the ${\bm k}$-integrals vanish by symmetry unless $j = i$. 
 Integrating by parts and evaluating the traces, one has 
\begin{align}
&
    {\rm  (f1) + \cdots + (f4)}
\nonumber \\ 
&=
   \frac{1}{2\pi}   (1+i\zeta)(-M{\mathcal R}^{-1}{\bm l}\, )^\beta 4A^z_i \, (-eE_i) 
\nonumber \\
& \times     \frac{1}{N} \sum_{\bm k}
    \Big[  (v^0_i)^2  \big\{   ((\mu^R)^2+3(T^R)^2-(J^R)^2)(\mu^AJ^R-J^A\mu^R)  \big\}
\nonumber \\&\quad\qquad\quad
    + (v'_i)^2  \big\{  ((\mu^R)^2+(T^R)^2-(J^R)^2)(\mu^AJ^R-J^A\mu^R)
\nonumber \\&\qquad\qquad\qquad\quad
        -2\mu^RT^R(T^RJ^A-J^RT^A)  -2T^RJ^R(T^R\mu^A-\mu^RT^A)
    \big\}
\nonumber \\&\quad\qquad\quad
    - 2v^0_i v'_i \,  \big\{ 2 \mu^RT^R(\mu^RJ^A-\mu^AJ^R)
        +  (  (\mu^R)^2-(J^R)^2+(T^R)^2  )  (T^RJ^A-J^RT^A) \big\}
    \Big] . 
\end{align}

Next we look at the fifth diagram in Fig.~\ref{fig:fig3} (f), 
\begin{align}
    {\rm (f5)}
&= \frac{1}{2\pi}  (1+i\zeta)  (-M{\mathcal R}^{-1}{\bm l})^\beta (-eE_i)
\nonumber \\&\qquad\times
    \frac{1}{N}\sum_{\bm k}
    {\rm tr}[  {\bm\sigma}^\perp\tau_3  G^R_{+} \sigma^\beta  G^R_{-}  (\partial_iv_{i-})  \tau_1  G^A_{-} ] , 
\end{align}
in which the wave vector ${\bm q}$ needs to be extracted from the Green's functions, 
$G^{R/A}_{\pm}\equiv G^{R/A}_{{\bm k}\pm {\bm q}/2}$, 
or from the velocity vertex, $v_{i\pm} = v_{i,{\bm k}\pm{\bm q}/2}$. 
 To first order in ${\bm q}$, and after integrating by parts, one has
\begin{align}
    {\rm (f5)}
&=  \frac{1}{2\pi}  (1+i\zeta) (-M{\mathcal R}^{-1}{\bm l})^\beta q_i (-eE_i) 
\nonumber \\
&\qquad\times
    \frac{1}{N} \sum_{\bm k}
    {\rm tr}[  {\bm\sigma}^\perp\tau_3 G^R  v_i  G^R  \sigma^\beta  G^R   v_i \tau_1 G^A ] . 
\end{align}
The trace is evaluated to give
\begin{align}
    {\rm (f5) }
&=  \frac{1}{2\pi} (1+i\zeta) \, 4\hat{z}\times (-M{\mathcal R}^{-1}{\bm l} \, ) \,  iq_i (-eE_i)
\nonumber \\  
& \times  \frac{1}{N}\sum_{\bm k}
    \bigg\{  (v^0_i )^2  ((\mu^R)^2+3(T^R)^2-(J^R)^2)  (J^R\mu^A-\mu^RJ^A)
\nonumber \\  
&\quad  + (v'_i)^2 \, 
        \Big[  ((\mu^R)^2+(T^R)^2-(J^R)^2)(\mu^AJ^R-\mu^RJ^A)
\nonumber \\
&\qquad 
        - 2\mu^RT^R(T^RJ^A-J^RT^A)   - 2T^RJ^R(T^R\mu^A-T^A\mu^R)  \Big]
\nonumber \\  
&\quad 
    -2v^0_iv'_i \,   \Big[  ((\mu^R)^2+(T^R)^2-(J^R)^2)(T^RJ^A-J^RT^A)
\nonumber \\ 
&\qquad\qquad 
        + 2\mu^RT^R(\mu^RJ^A-J^R\mu^A)  \Big]
    \bigg\} . 
\end{align}
 Taking $iq_i \to \partial_i$ 
and using the relation
\begin{align}
    {\hat z}\times(\partial_i{\mathcal R}^{-1}) \, {\bm l}
&=  {\hat z}\times({\bm A}_i\times{\mathcal R}^{-1}{\bm l}\, )
\\
&= -{A}^z_i  {\mathcal R}^{-1}{\bm l}  , 
\end{align}
we see that (f5) cancels the gauge-noninvariant terms in ${\rm  (f1) + \cdots + (f4)}$.

 The vertex correction is taken into account by the replacement, 
\begin{align}
    \sigma^\perp\tau_3 &\to    i \sigma^\perp\sigma^3  \Lambda_3  , 
\end{align}
where $\Lambda_3$ is given by
\begin{align}
    \Lambda_3
=  \frac{2 \pi n_{\rm i}u_{\rm i}^2}{1-\Lambda_1}  \,
    \frac{1}{N} \sum_{\bm k}
    \frac{(\mu-\varepsilon'_{\bm k})\gamma_3+M\gamma_0 }{|(\mu-\varepsilon'_{\bm k})\gamma_0 + M\gamma_3|}
    \delta((\mu-\varepsilon'_{\bm k})^2-E_{\bm k}^2)  . 
\end{align}
 This is simplified in the AF transport limit, 
\begin{align}
    \Lambda_3 =    \frac{\gamma_0}{M}  \qquad (t' = 0) , 
\end{align}
and also in the FM transport limit, 
\begin{align}
    \Lambda_3 =   \frac{\gamma_0}{M}  
\qquad (t = 0)  .
\end{align}

The next four diagrams (with vertex corrections) give
\begin{align}
    {\rm (f1V) + \cdots + (f4V)}
&=  \frac{1}{2\pi}   i  \Lambda_3  (1+i\zeta)(-M{\mathcal R}^{-1}{\bm l}\, )^\beta (-eE_i)
\nonumber \\
& \times     \frac{1}{N}  \sum_{\bm k}
    \Big\{  {\rm tr}[ \sigma^\perp\sigma^3 G^R \sigma^\beta G^R  A_i  v_i G^R v_i  G^A ]
\nonumber \\
&\qquad\qquad +
    {\rm tr}[ \sigma^\perp\sigma^3 G^R A_i  v_i G^R \sigma^\beta G^R v_i G^A ]
\nonumber \\
&\qquad\qquad +  {\rm tr}[  \sigma^\perp\sigma^3 G^R \sigma^\beta  G^R v_i G^A  A_i v_i  G^A  ]
\nonumber \\
&\qquad\qquad +
    {\rm tr}[ \sigma^\perp\sigma^3 G^R  \sigma^\beta G^R A_i  (\partial_iv_i) G^A  ]   \Big\}
\\
&= \frac{1}{2\pi}   i \Lambda_3  (1+i\zeta)(-M{\mathcal R}^{-1}{\bm l}\, ) \, 4A^z_i (-eE_i)
\nonumber \\
& \times  \frac{1}{N} \sum_{\bm k}
    \bigg\{  (v^0_i)^2  \Big[  ((\mu^R)^2+(T^R)^2-(J^R)^2)(\mu^R\mu^A+T^RT^A-J^RJ^A)
\nonumber \\
&\qquad\qquad 
       + 2\mu^RT^R(T^R\mu^A+\mu^RT^A)  - 2T^RJ^R(J^RT^A+T^RJ^A) \Big]
\nonumber \\
&\qquad +  (v'_i)^2 \Big[ ((\mu^R)^2 - (T^R)^2 - (J^R)^2)(\mu^R\mu^A-T^RT^A-J^RJ^A)
\nonumber \\
&\qquad\qquad\quad + 4\mu^RT^R(\mu^AT^R+\mu^RT^A)  \Big] 
\nonumber \\
&\qquad + 2v^0_i v'_i \,   \Big[  ((\mu^R)^2+(T^R)^2-(J^R)^2)(T^R\mu^A+\mu^RT^A)
\nonumber \\
&\qquad\qquad\quad + 2\mu^RT^R(\mu^R\mu^A+T^RT^A-J^RJ^A)
    \Big] 
    \bigg\} . 
\end{align}

 The last diagram in Fig. \ref{fig:fig3} (f) gives 
\begin{align}
    {\rm (f5V)}
&=  \frac{1}{2\pi}  i \Lambda_3  (1+i\zeta)(-M{\mathcal R}^{-1}{\bm l}\, )^\beta q_i \, (-eE_i) 
\nonumber \\
&\qquad \times \frac{1}{N}  \sum_{\bm k}  {\rm tr}[ \sigma^\perp\sigma^3
        G^R  v_i  G^R  \sigma^\beta G^R   v_i  G^A ]  
\\
&=   \frac{1}{2\pi}   i  \Lambda_3 
    (1+i\zeta)(-M{\mathcal R}^{-1} {\bm l} \, ) \, q_i \, (-eE_i) \, 
    {\rm tr}[ \sigma^\perp  \sigma^3 \sigma^\beta ] 
\nonumber \\
& \times   \frac{1}{N} \sum_{\bm k}
    \bigg\{ 
    (v^0_i)^2  \Big[  ((\mu^R)^2+(T^R)^2-(J^R)^2)(\mu^R\mu^A+T^RT^A-J^RJ^A)
\nonumber \\
&\qquad\qquad\quad + 2\mu^RT^R(T^R\mu^A+\mu^RT^A) - 2T^RJ^R(J^RT^A+T^RJ^A) \Big]
\nonumber \\
&\qquad 
  + (v'_i)^2 \Big[  ((\mu^R)^2-(T^R)^2-(J^R)^2)(\mu^R\mu^A-T^RT^A-J^RJ^A)
\nonumber \\
&\qquad\qquad\quad 
  + 4 \mu^R T^R (\mu^AT^R + \mu^RT^A)  \Big]
\nonumber \\
&\qquad + 2v^0_iv'_i \,  \Big[  ((\mu^R)^2+(T^R)^2-(J^R)^2)(\mu^RT^A+T^R\mu^A)
\nonumber \\
&\qquad\qquad\quad + 2\mu^RT^R(\mu^R\mu^A+T^RT^A-J^RJ^A)
    \Big] 
    \bigg\} . 
\end{align}

Thus, we see that the gauge-noninvariant terms in (f5V) are canceled out by (f1V) $+\cdots$ (f4V).
It is clear that the effect of vertex correction is absent when $t=0$. 
When $t'=0$, (f5V) is equal to ${\rm (f5)} \times (\mu^2-M^2)/(2M^2)$.
Adding up all diagrams in Fig. \ref{fig:fig3} (f), 
the current induced staggered spin density is given by

\begin{align}
\langle {\bm\sigma}_n\rangle
&=
    \frac{2}{\pi}  \, (-M{\bm n}\times\partial_i {\bm l} \, ) \,  (-eE_i)
\, \frac{1}{N}\sum_{\bm k}
    \bigg\{  (v^0_i )^2 C_1
  + (v'_i)^2 \,  C_2
    +2v^0_iv'_i \,  C_3 
    \bigg\}  
\end{align}
where the coefficient of $(v^0_i)^2$ is given by
\begin{align}
C_1=&
    (1+i\zeta)
 \Big[((\mu^R)^2+3(T^R)^2-(J^R)^2)  (J^R\mu^A-\mu^RJ^A)
    \nonumber \\&\qquad
    +i\Lambda_3\Big\{  ((\mu^R)^2+(T^R)^2-(J^R)^2)(\mu^R\mu^A+T^RT^A-J^RJ^A)
\nonumber \\
&\qquad\qquad + 2\mu^RT^R(T^R\mu^A+\mu^RT^A) - 2T^RJ^R(J^RT^A+T^RJ^A) \Big\}
    \Big]
 +{\rm c.c.} 
\end{align}
 the coefficient of $(v'_i)^2$ is given by
\begin{align}
C_2 =&
    (1+i\zeta)
    \Big[  ((\mu^R)^2+(T^R)^2-(J^R)^2)(\mu^AJ^R-\mu^RJ^A)
\nonumber \\
&\qquad 
        - 2\mu^RT^R(T^RJ^A-J^RT^A)   - 2T^RJ^R(T^R\mu^A-T^A\mu^R)  
    \nonumber \\&\qquad
    +i\Lambda_3\Big\{  
     ((\mu^R)^2-(T^R)^2-(J^R)^2)(\mu^R\mu^A-T^RT^A-J^RJ^A)
\nonumber \\
&\qquad\qquad  
      + 4 \mu^R T^R (\mu^AT^R + \mu^RT^A)
    \Big\}
    \Big]
 +{\rm c.c.} 
\end{align}
and the coefficient of $2v^0_iv'_i$ is given by
\begin{align}
C_3 =&
(1+i\zeta)
\Big[  -((\mu^R)^2+(T^R)^2-(J^R)^2)(T^RJ^A-J^RT^A)
\nonumber \\ 
&\qquad
        - 2\mu^RT^R(\mu^RJ^A-J^R\mu^A)  
    \nonumber \\&\qquad
    +i\Lambda_3 \Big\{  ((\mu^R)^2+(T^R)^2-(J^R)^2)(\mu^RT^A+T^R\mu^A)
\nonumber \\
&\qquad\qquad + 2\mu^RT^R(\mu^R\mu^A+T^RT^A-J^RJ^A)
    \Big\}
    \Big]
 +{\rm c.c.} 
\end{align}


\bibliographystyle{jpsj}
\bibliography{ref}